\documentclass[12pt,a4paper]{article}
\usepackage{fullpage}
\usepackage{float}
\usepackage{floatflt}
\usepackage[centertags]{amsmath}
\allowdisplaybreaks[1]
\numberwithin{equation}{section}
\usepackage{amsbsy}
\usepackage{amsfonts}
\usepackage{amssymb}
\usepackage[dvips]{graphicx}
\usepackage[dvips]{hyperref}
\usepackage[square,comma,numbers]{natbib}
\usepackage{tocbibind}
\usepackage[thinspace,derived,squaren]{SIunits}
\begin{document}

\begin{flushright}
\textsf{30 October 2003}
\\
\textsf{hep-ph/0302026}
\end{flushright}

\vspace{1cm}

\begin{center}
\large
\textbf{Coherence and Wave Packets in Neutrino Oscillations}
\normalsize
\\[0.5cm]
\large
Carlo Giunti
\normalsize
\\[0.5cm]
INFN, Sezione di Torino, and Dipartimento di Fisica Teorica,
\\
Universit\`a di Torino,
Via P. Giuria 1, I--10125 Torino, Italy
\\[0.5cm]
\begin{minipage}[t]{0.8\textwidth}
\begin{center}
\textbf{Abstract}
\end{center}
General arguments in favor
of the necessity of a wave packet description of
neutrino oscillations are presented,
drawing from analogies with other wave phenomena.
We present a wave packet description
of neutrino oscillations in stationary beams
using the density matrix formalism.
Recent claims of the necessity of
an equal energy of different massive neutrinos are refuted.
\end{minipage}
\end{center}

\newpage
\tableofcontents
\newpage

\section{Introduction}
\label{Introduction}

The physics of massive and mixed neutrinos
is one of the hot topics in today's research in
high energy physics,
following the discovery of oscillations
of atmospheric and solar neutrinos
(see the reviews in Refs.~\cite{Bilenkii:2001yh,%
Gonzalez-Garcia:2002dz,%
Kayser:2002qs,%
Bilenky:2002aw}
and references therein,
as well as the references in
\cite{Neutrino-Unbound}).

Neutrino oscillations have been proposed in the late 50's
by Pontecorvo
\cite{Pontecorvo:1957cp,Pontecorvo-58}.
The oscillations are generated by the interference
of different massive neutrinos,
which are produced and detected coherently
because of their very small mass difference.

In 1962 Maki, Nakagawa and Sakata
\cite{Maki:1962mu}
considered for the first time a model
with mixing of different neutrino flavors.
In 1967
Pontecorvo
proposed the possibility of solar neutrino oscillations
\cite{Pontecorvo:1968fh},
before the discovery in 1968 of the solar neutrino problem
in the Homestake experiment
\cite{Davis:1968cp}.
In 1969
Pontecorvo and Gribov
considered $\nu_e\to\nu_\mu$ oscillations
as a possible explanation of the solar neutrino problem
\cite{Gribov:1969kq}.

The theory of neutrino oscillations
in the plane-wave approximation
was developed in the middle 70's
by Eliezer and Swift
\cite{Eliezer:1976ja},
Fritzsch and Minkowski
\cite{Fritzsch:1976rz},
Bilenky and Pontecorvo
\cite{Bilenky:1976yj},
and beautifully reviewed by Bilenky and Pontecorvo
in Ref.~\cite{Bilenky:1978nj}.

In 1976 Nussinov
\cite{Nussinov:1976uw}
for the first time
considered the wave packet nature of propagating neutrinos
and inferred the existence of a coherence length,
beyond which the interference of
different massive neutrinos
is not observable.
This is due to the different group velocities
of different massive neutrinos,
that causes a separation of their wave packets.
In 1996
Kiers, Nussinov and Weiss
\cite{Kiers:1996zj}
first pointed out the importance
of the detection process for the coherence of
neutrino oscillations
and discussed some implications
for the wave packet approach.

In 1981 Kayser
\cite{Kayser:1981ye}
presented the first detailed discussion
of the quantum mechanical problems
of neutrino oscillations,
pointing out the necessity of a wave packet treatment.

Wave packet models
of neutrino oscillations have been later developed
in the framework of quantum mechanics
\cite{Giunti-Kim-Lee-Whendo-91,%
Giunti:1992sx,%
Dolgov-Morozov-Okun-Shchepkin-97,%
Giunti-Kim-Coherence-98,%
Dolgov:1999sp,%
Dolgov:2002wy,%
hep-ph/0202063}
and
in the framework of quantum field theory
\cite{Giunti-Kim-Lee-Lee-93,%
Giunti-Kim-Lee-Whendo-98,%
Kiers-Weiss-PRD57-98,%
Cardall-Coherence-99,%
Beuthe:2002ej,%
hep-ph/0205014}
(see also Ref.~\cite{CWKim-book}
and the reviews in
Refs.~\cite{Zralek-oscillations-98,Beuthe:2001rc}).

In spite of the well-known fact that
in quantum theory localized particles are described by
wave packets
and in spite of the success of the wave packet treatment
of neutrino oscillations,
some authors have presented arguments
against such approach.
In Section~\ref{No source of waves vibrates indefinitely}
we briefly review well-known evidences
of the wave packet nature of light,
which, by analogy,
imply a wave packet nature of neutrinos.
In Sections~\ref{Bound states}
and \ref{Stationary beams}
we present objections to arguments
against a wave packet description of neutrinos.
In Section~\ref{Stationary beams}
we present also a wave packet derivation of neutrino oscillations
in the density matrix formalism
and we discuss some implications for the case of stationary beams.

Another debated problem
in the theory of neutrino oscillations
is the determination of the energies of massive neutrinos.
In Section~\ref{Energy of massive neutrinos}
we present objections to
recently proposed arguments
in favor of the equality
of the energies of different massive neutrinos.

\section{No source of waves vibrates indefinitely}
\label{No source of waves vibrates indefinitely}

The title of this section is taken from the beginning sentence
of Section~11.11
of Ref.~\cite{Jenkins-White-FundamentalsOfOptics-1981},
which concludes with the following paragraph:
\begin{quote}
In light sources,
the radiating atoms emit wave trains of finite length.
Usually,
because of collisions or damping arising from other causes,
these packets are very short.
According to the theorem mentioned above\footnote{
``The largest the number $N$ of waves in the group,
the smaller the spread $\Delta\lambda$,
and in fact theory shows that
$\Delta\lambda/\lambda_0$
is approximately equal to $1/N$.''
(Section~11.11
of Ref.~\cite{Jenkins-White-FundamentalsOfOptics-1981}.)
},
the consequence is that the spectrum lines will not be very narrow
but will have an appreciable width $\Delta\lambda$.
A measurement of this width will yield the effective
``lifetime''
of the electromagnetic oscillators
in the atoms and the average length of the wave packets.
A low-pressure discharge through the vapor of mercury
containing the single isotope
$^{198}\mathrm{Hg}$
yields very sharp spectral lines, of width about
$\unit{0.005}{\angstrom}$.
Taking the wavelength of one of the brightest lines,
$\unit{5461}{\angstrom}$,
we may estimate that there are roughly $10^6$
waves in a packet
and that the packets themselves are some 50 cm long.
\end{quote}

The broadening of optical lines
due to the finite lifetime of atomic transitions
is known as \emph{natural linewidth}.

It is interesting to note that the broadening
of optical lines
was well known to experimental physicists
in the nineteen century
and explained by classical models
before the
advent of quantum theory
(see Ref.~\cite{Breene-57} and references therein).
In 1895 Michelson
\cite{Michelson-1895}
listed among the hypotheses formulated before that time
to account for line broadening
\begin{quote}
3. The exponential diminution in amplitude
of the vibrations due to communication of energy
to the surrounding medium or to other causes.
\end{quote}
As explained in \cite{Breene-57},
\begin{quote}
We consider an emitting atom
which we shall proceed to remove to infinity and reduce the ``temperature''
to the point where, classically at least,
no translational motion exists.
Now from the classical picture of a vibrating electron
or the simple picture of a pair of energy levels
between which our radiation transition takes place,
we should expect these conditions
to yield a spectral line of a single frequency.
We, of course, do not obtain this result, but,
rather, we obtain the familiar natural line shape
which is attributable to Michelson's Cause 3.
\end{quote}
A classical derivation of the natural linewidth
can be found in Section 17.7 of Ref.~\cite{Jackson-book-75}.
Let us emphasize that the natural linewidth of atomic lines
has been \emph{observed experimentally}
(see Section 21.4 of Ref.~\cite{White-book-34})!

Another important cause of line broadening known in 1895 was
\cite{Michelson-1895}
\begin{quote}
4.
The change in wavelength due to the Doppler effect
of the component of the velocity of the vibrating atom
in the line of sight.
\end{quote}
The Doppler line broadening due to the thermal motion
of atoms in a medium was calculated by Rayleigh in 1889
\cite{Rayleigh-1889}.
This important effect,
that must be aways taken into account
in calculating the spectral shape
of monochromatic beams,
does not concern us here,
because it does not generate a coherent
broadening.
It is simply due to the different motion of
different atoms,
whose radiation is incoherent. 

To these (and other) causes of line broadening
Michelson added in 1895
\cite{Michelson-1895}
\begin{quote}
5.
The limitation of the number of regular vibrations
by more or less abrupt changes of phase amplitude
or
plane of vibration caused by collisions.
\end{quote}
This important coherent effect has been called with several names,
among which
\emph{collision broadening},
\emph{pressure broadening}
and
\emph{interruption broadening},
it has been studied in depth by many authors
(see, for example, Refs.\cite{Breene-57,%
Loudon-TheQuantumTheoryOfLight-92})
and it has been \emph{observed experimentally}
(see Section 21.5 of Ref.~\cite{White-book-34})!

In quantum theory,
the fact that ``no source of waves vibrates indefinitely''
implies that all particles are produced as wave packets,
whose size is determined by the finite lifetime of
the parent particle,
or by its finite mean free path if the production process
occurs in a medium.
Since also
no wave detector vibrates indefinitely,
it is clear that all particles are also
detected as wave packets.

Usually, at least in high energy physics,
the wave packet character of particles is not important
and they can be well approximated by plane waves.
The phenomenon of neutrino oscillations is an exception,
as shown in 1981 by Kayser
\cite{Kayser:1981ye},
because the localization of the source and detector
requires a wave packet treatment.

The wave packet approach to neutrino oscillations
implemented in
Refs.~\cite{Giunti-Kim-Lee-Whendo-91,%
Giunti:1992sx,%
Dolgov-Morozov-Okun-Shchepkin-97,%
Giunti-Kim-Coherence-98,%
Dolgov:1999sp,%
hep-ph/0202063}
in the framework of quantum mechanics
and in
Refs.~\cite{Giunti-Kim-Lee-Lee-93,%
Giunti-Kim-Lee-Whendo-98,%
Kiers-Weiss-PRD57-98,%
Cardall-Coherence-99,%
Beuthe:2002ej,%
hep-ph/0205014}
in the framework of quantum field theory
allows to
derive the neutrino oscillation
probability from first principles in a consistent framework.

In the light of these considerations,
it seems rather surprising that some physicists
do not agree with a wave packet description of neutrinos
and put forward arguments
in favor of a unique value of neutrino energy.
A tentative explanation of this approach stems from the
common education and practice in physics
of working with energy eigenstates,
forgetting that they are only approximations
of the states of physical systems in the real world\footnote{
I would like to thank I. Pesando for an illuminating
discussion about this point.
}.
In the following Sections
we present a critical discussion
of recently advocated approaches to neutrino oscillations
in which neutrinos are not described
by wave packets with some spread in momentum and energy.

\section{Bound states}
\label{Bound states}

Bound states of atomic electrons are usually calculated
with different degrees of approximation
taking into account the static potential generated
by the nucleus
and other electrons
and
neglecting the coupling of the electron
with the electromagnetic radiation field.
In this approximation,
bound states are stationary and have definite energy.
In spite of the obvious fact that
in such approximation
electrons
cannot jump from one bound state to another,
the energy difference between two stationary bound states
is sometimes associated
without any further explanation
with the energy of the photon
emitted or absorbed in the transition of an electron
from one bound state to the other.

However,
in order to describe the transitions
between bound states it is necessary
to take into account the coupling of the electron
with the time-dependent electromagnetic radiation field.
Since the bound states are not eigenstates of the
full Hamiltonian,
in reality they are not stationary and
they do not have definite energy.
The lack of a definite energy is a necessary
requirement for the existence of transitions:
since
the Schr\"odinger equation
implies that the time evolution of
energy eigenstates evolve by a phase,
orthogonality between different
energy eigenstates is conserved in time,
excluding transitions.

The fact that unstable systems
do not have a definite energy
is well-known to high energy physicists
from their experience with resonances
(hadronic excited states)
and highly unstable particles
($Z$ and $W$ bosons),
which do not have a definite mass
(energy in the rest frame).
Their mass width is inversely proportional
to the decay rate.

From these considerations
it follows that the quantum field theoretical
model of neutrino oscillations in the process
\begin{equation}
n \to p + e^- + \bar\nu_e
\xrightarrow{\bar\nu_e \to \bar\nu_e}
\bar\nu_e + e^- \to \bar\nu_e + e^-
\label{GS01}
\end{equation}
presented
in Ref.~\cite{Grimus-Stockinger-96},
in which a neutrino is assumed to be produced by a nuclear
$\beta$-decay between two stationary bound states 
with definite energy
and detected through scattering with an electron
in a stationary atomic state with definite energy,
is unrealistic,
not to mention the description of the final-state
electrons by plane waves, which is
in contradiction with
the fact that the electrons interact with the surrounding medium
and with
the necessity to observe
the final lepton in the detection process,
as already noted in \cite{hep-ph/0205014}.
Of course,
once the matters of principle are clear,
one can consider the assumptions
in Ref.~\cite{Grimus-Stockinger-96}
as approximations acceptable in some cases.
One must also notice that
Ref.~\cite{Grimus-Stockinger-96}
is very interesting for the useful theorem
proved in the appendix,
which has been used in the quantum field theoretical
calculation of neutrino oscillations in the wave packet approach
in Refs.~\cite{Giunti-Kim-Lee-Whendo-98,Beuthe:2002ej}.

Developing the technique
proposed in Ref.~\cite{Grimus-Stockinger-96},
the authors of
Refs.~\cite{Grimus-Mohanty-Stockinger-98,Grimus-Mohanty-Stockinger-99}
considered neutrino oscillations in the process
\begin{equation}
\mu^+ \to e^+ + \nu_e + \bar\nu_\mu
\xrightarrow{\bar\nu_\mu \to \bar\nu_e}
\bar\nu_e + p \to n + e^+
\label{GS02}
\end{equation}
with the muon correctly described by a wave packet,
but the proton was still described by an unrealistic
stationary bound state with definite energy.
Furthermore,
all final particles were described by plane waves,
in contradiction with the fact that
the positrons and neutron
interact with the surrounding medium and
the necessity to observe the final positron
in the detection process
in order to measure oscillations.
Localized interactions of a particle with a detector
obviously imply that the particle cannot be described by an
unlocalized plane wave
and a wave packet description is necessary.

\section{Stationary beams}
\label{Stationary beams}

The author of the very interesting Ref.~\cite{Stodolsky-unnecessary-98}
wrote:
\begin{quote}
\ldots
many wavepacket discussions for the coherence
properties of particle beams
are unnecessary since they deal with stationary sources;
and when the problem is stationary,
essentially all information is in the energy spectrum.
\end{quote}
Indeed, as proved in Ref.~\cite{Kiers:1996zj},
\begin{quote}
\ldots
under very general assumptions it is not possible to distinguish experimentally neutrinos
produced in some region of space as wave packets
from those produced in the same region of space as
plane waves with the same energy distribution.
\end{quote}

It is not clear if the purpose of
Ref.~\cite{Stodolsky-unnecessary-98}
was to show that neutrinos are not described by wave packets,
but it is certain that it has been interpreted in this way by some
physicists.
However,
Ref.~\cite{Stodolsky-unnecessary-98}
certainly does not provide any help
for the derivation of neutrino oscillations
in the framework of quantum field theory.
Even in the framework of quantum mechanics
Ref.~\cite{Stodolsky-unnecessary-98}
does not go beyond the well-known
discussion of neutrino oscillations
assuming equal energies
for the different massive neutrino components
\cite{Lipkin-nonexperiments-95,Grossman-Lipkin-spatial-97},
which has been shown to be unrealistic
in Refs.~\cite{Giunti:2000kw,Giunti:2001kj}
and is further criticized in the Section~\ref{Energy of massive neutrinos}.

An interpretation of
Ref.~\cite{Stodolsky-unnecessary-98}
as a proof that neutrinos are not described by wave packets
seems to stem from a confusion between microscopic and macroscopic
stationarity,
as already remarked in \cite{Cardall-Coherence-99,Giunti:2000kw}.
Since most neutrino sources
emit neutrinos
at a constant rate over macroscopic intervals of time,
they can be considered macroscopically stationary.
However,
the microscopic processes of neutrino production
(nuclear or particle decay, etc.)
are certainly not stationary
and the single neutrino does not know if
it will be part of a stationary beam or not.
Because of the localization in space and time of the production and detection processes,
a description of neutrino oscillations
in terms of wave packets
is inescapable.

Of course,
after the matters of principle are settled,
neutrino oscillations are
derived from first principles in a consistent theoretical framework,
and an estimation of the coherence length
has shown that it is much longer than
the source-detector distance in a neutrino oscillation experiment,
the analysis of the experimental data
can be performed using the standard
oscillation probability
obtained in the plane wave approximation.
This turns out to be the case in all present-day
neutrino oscillation experiments.

In the following part of this section we present
a wave packet derivation of neutrino oscillations in the
density matrix formalism,
which is suitable for the description of a stationary beam,
as emphasized in Ref.~\cite{Stodolsky-unnecessary-98}.
Our calculation follows
a method similar to the one presented in Ref.~\cite{Giunti:1992sx}.

In a quantum-mechanical wave packet treatment of neutrino oscillations
a neutrino produced at the origin of the space-time coordinates
by a weak interaction process with definite flavor $\alpha$
($\alpha=e,\mu,\tau$)
and propagating along the $x$ axis
is described by the state
\begin{equation}
|\nu_\alpha(x,t)\rangle
=
\sum_k U_{\alpha k}^* \psi_k(x,t) |\nu_k\rangle
\,,
\label{001}
\end{equation}
where $U$ is the mixing matrix,
$|\nu_k\rangle$
is the state of a neutrino with mass $m_k$
and
$\psi_k(x,t)$
is its wave function.
Approximating the momentum distribution
of the massive neutrino $\nu_k$
with a gaussian,
\begin{equation}
\psi_k(p)
=
\left( 2 \pi {\sigma_p^{\mathrm{P}}}^2 \right)^{-1/4}
\exp\left[
-
\frac{ \left( p - p_k \right)^2 }{ 4 {\sigma_p^{\mathrm{P}}}^2 }
\right]
\,,
\label{002}
\end{equation}
where
$p_k$ is the average momentum and $\sigma_p^{\mathrm{P}}$
is the momentum uncertainty determined by the production process,
the wave function is
\begin{equation}
\psi_k(x,t)
=
\int \frac{ \mathrm{d}p }{ \sqrt{2\pi} }
\psi_k(p)
e^{ipx-iE_k(p)t}
\,,
\label{004}
\end{equation}
with the energy $E_k(p)$ given by\footnote{
We do not make unnecessary assumptions about
the values of the energies and momenta of different massive neutrinos.
The claim that different massive neutrinos
must have the same energy has been confuted in
Refs.~\cite{Giunti:2000kw,Giunti:2001kj}
and in further discussed
in Section~\ref{Energy of massive neutrinos}.
}
\begin{equation}
E_k(p) = \sqrt{ p^2 + m_k^2 }
\,.
\label{005}
\end{equation}
If the gaussian momentum distribution
(\ref{002})
is sharply peaked around the mean momentum $p_k$
(\textit{i.e.} if $\sigma_p^{\mathrm{P}} \ll E_k^2(p_k)/m_k$),
the energy $E_k(p)$ can be approximated by
\begin{equation}
E_k(p)
\simeq
E_k
+
v_k \left( p - p_k \right)
\,,
\label{006}
\end{equation}
where
\begin{equation}
E_k
\equiv
E_k(p_k)
=
\sqrt{ p_k^2 + m_k^2 }
\label{007}
\end{equation}
is the average energy
and
\begin{equation}
v_k
\equiv
\left.
\frac{ \partial E_k(p) }{ \partial p }
\right|_{p=p_k}
=
\frac{ p_k }{ E_k }
\,.
\label{008}
\end{equation}
Under this approximation,
the integration over $p$ in Eq.~(\ref{004})
is gaussian and leads to the wave packet in coordinate space
\begin{equation}
\psi_k(x,t)
=
\left( 2 \pi {\sigma_x^{\mathrm{P}}}^2 \right)^{-1/4}
\exp\left[
- i E_k t + i p_k x
-
\frac{ \left( x - v_k t \right)^2 }{ 4 {\sigma_x^{\mathrm{P}}}^2 }
\right]
\,,
\label{009}
\end{equation}
where
\begin{equation}
\sigma_x^{\mathrm{P}}
=
\frac{ 1 }{ 2 \sigma_p^{\mathrm{P}} }
\label{010}
\end{equation}
is the spatial width of the wave packet.
From Eq.~(\ref{009})
one can see that $v_k$
is the group velocity
of the wave packet of the massive neutrino $\nu_k$.

The pure state in Eq.~(\ref{001})
can be written in the form of a density matrix operator as
\begin{equation}
\hat\rho_\alpha(x,t)
=
|\nu_\alpha(x,t)\rangle
\langle\nu_\alpha(x,t)|
\,.
\label{011}
\end{equation}
Using Eq.~(\ref{009}),
we have
\begin{eqnarray}
&&
\hat\rho_\alpha(x,t)
=
\frac{ 1 }{ \sqrt{ 2 \pi {\sigma_x^{\mathrm{P}}}^2 } }
\sum_{k,j}
U_{\alpha k}^* U_{\alpha j}
\exp\left[
- i \left( E_k - E_j \right) t + i \left( p_k - p_j \right) x
\vphantom{
\frac{ \left( x - v_k t \right)^2 }{ 4 {\sigma_x^{\mathrm{P}}}^2 }
}
\right.
\nonumber
\\
&&
\hspace{6.5cm}
\left.
-
\frac{ \left( x - v_k t \right)^2 }{ 4 {\sigma_x^{\mathrm{P}}}^2 }
-
\frac{ \left( x - v_j t \right)^2 }{ 4 {\sigma_x^{\mathrm{P}}}^2 }
\right]
|\nu_k\rangle
\langle\nu_j|
\,.
\label{012}
\end{eqnarray}
This space and time dependent density matrix operator
describes neutrino oscillations in space and time.
Although
in laboratory experiments
it is possible to measure neutrino oscillations in time
through the measurement of both the
production and detection processes
(see Ref.~\cite{Okun:2000-SNEGIRI})\footnote{
Let us also mention the possibility of
time-dependent flavor oscillations of neutrinos
in the primordial cosmological plasma
(see Ref.~\cite{Dolgov:2002wy} and references therein).
},
in all existing neutrino oscillation experiments
only the source-detector distance is known.
In this case the relevant density matrix operator
$\hat\rho_\alpha(x)$
is given by the time average of $\hat\rho_\alpha(x,t)$.
Since the integral over time is gaussian,
one easily obtains\footnote{
The density matrix operator in Eq.~(\ref{013})
is normalized by
$
\sum_{\alpha}
\hat\rho_\alpha(x)
=
\hat{I}
$,
where $\hat{I}$
is the identity operator.
}
\begin{eqnarray}
&&
\hat\rho_\alpha(x)
=
\sum_{k,j}
U_{\alpha k}^* U_{\alpha j}
\exp\left\{
- i
\left[
\frac{v_k+v_j}{v_k^2+v_j^2} \left( E_k - E_j \right)
- \left( p_k - p_j \right)
\right]
x
\right.
\nonumber
\\
&&
\hspace{4.5cm}
\left.
-
\frac{ \left( v_k - v_j \right)^2 x^2 }{ 4 \left( v_k^2 + v_j^2 \right) {\sigma_x^{\mathrm{P}}}^2 }
- \frac{ \left( E_k - E_j \right)^2 }{ 4 \left( v_k^2 + v_j^2 \right) {\sigma_p^{\mathrm{P}}}^2 }
\right\}
|\nu_k\rangle
\langle\nu_j|
\,.
\label{013}
\end{eqnarray}
Since this density matrix operator is time independent,
it is suitable for the description of a stationary
beam in neutrino oscillation experiments,
as noticed in Ref.~\cite{Stodolsky-unnecessary-98}.

In order
to obtain the oscillation probability,
it is convenient to simplify Eq.~(\ref{013})
for the realistic case of extremely relativistic neutrinos
(see the discussion in Section~2 of Ref.~\cite{hep-ph/0205014}).
In general,
the average massive neutrino energies $E_k$
can be written as
\begin{equation}
E_k \simeq E + \xi_{\mathrm{P}} \, \frac{m_k^2}{2E}
\,,
\label{101}
\end{equation}
where $E$ is the neutrino energy in the limit of zero mass
and $\xi_{\mathrm{P}}$ is a dimensionless quantity
that depends from the characteristics of the production process\footnote{
\label{xip}
The quantity
$\xi_{\mathrm{P}}$
cannot be calculated in a quantum mechanical
framework,
but its value can be estimated from energy-momentum conservation
in the production process.
The calculation of the value of
$\xi_{\mathrm{P}}$
requires a quantum field theoretical treatment
(see Ref.~\cite{hep-ph/0205014}).
}.
From Eq.~(\ref{005}),
the corresponding momentum in the relativistic approximation is
\begin{equation}
p_k \simeq E - \left( 1 - \xi_{\mathrm{P}} \right) \frac{m_k^2}{2E}
\,.
\label{102}
\end{equation}
The generality of these relations can be understood by noting that,
since the neutrino mass
$m_k$
enters quadratically in the energy-momentum dispersion relation
(\ref{005}),
its
the first order contribution
to the energy and momentum
must be proportional
to $m_k^2$ and
inversely proportional to $E$,
which is the only available quantity with dimension of energy.

Using Eqs.~(\ref{101}), (\ref{102}) and the relativistic approximation
\begin{equation}
v_k
\simeq
1 - \frac{ m_k^2 }{ 2 E_k^2 }
\,,
\label{103}
\end{equation}
we obtain
\begin{equation}
\hat\rho_\alpha(x)
=
\sum_{k,j}
U_{\alpha k}^* U_{\alpha j}
\exp\left[
- i
\frac{ \Delta{m}^2_{kj} x }{ 2 E }
-
\left(
\frac{ \Delta{m}^2_{kj} x }{ 4 \sqrt{2} E^2 \sigma_x^{\mathrm{P}} }
\right)^2
-
\left(
\xi_{\mathrm{P}}
\frac{ \Delta{m}^2_{kj} }{ 4 \sqrt{2} E \sigma_p^{\mathrm{P}} }
\right)^2
\right]
|\nu_k\rangle
\langle\nu_j|
\,,
\label{014}
\end{equation}
with
$ \Delta{m}^2_{kj} \equiv m_k^2 - m_j^2 $.

In analogy with the production process,
we describe the process of detection of a
neutrino with flavor $\beta$ at the coordinate $x=L$ with the
operator
\begin{align}
\hat{\mathcal{O}}_\beta&(x-L)
=
\sum_{k,j}
U_{\beta k}^* U_{\beta j}
\nonumber
\\
&
\times
\exp\left[
- i
\frac{ \Delta{m}^2_{kj} \left( x - L \right) }{ 2 E }
-
\left(
\frac{ \Delta{m}^2_{kj} \left( x - L \right) }{ 4 \sqrt{2} E^2 \sigma_x^{\mathrm{D}} }
\right)^2
-
\left(
\xi_{\mathrm{D}}
\frac{ \Delta{m}^2_{kj} }{ 4 \sqrt{2} E \sigma_p^{\mathrm{D}} }
\right)^2
\right]
|\nu_k\rangle
\langle\nu_j|
\,,
\label{015}
\end{align}
where
$\xi_{\mathrm{D}}$
is a dimensionless quantity
that depends from the characteristics of the detection process
(see footnote~\ref{xip}),
$\sigma_p^{\mathrm{D}}$
is the momentum uncertainty of the detection process
and
$ \sigma_x^{\mathrm{D}} = 1 / 2 \sigma_p^{\mathrm{D}} $.

The probability of
$ \nu_\alpha \to \nu_\beta $
transitions
is given by
\begin{equation}
P_{\nu_\alpha\to\nu_\beta}(L)
=
\mathrm{Tr}\left( \hat\rho_\alpha(x) \hat{\mathcal{O}}_\beta(x-L) \right)
=
\int \mathrm{d}x
\sum_k \langle\nu_k| \hat\rho_\alpha(x) \hat{\mathcal{O}}_\beta(x-L) |\nu_k\rangle
\,,
\label{016}
\end{equation}
which leads to
\begin{equation}
P_{\nu_\alpha\to\nu_\beta}(L)
=
\sum_{k,j}
U_{\alpha k}^* U_{\alpha j} U_{\beta k} U_{\beta j}^*
\exp\left[
- 2 \pi i \frac{ L }{ L^{\mathrm{osc}}_{kj} }
-
\left(
\frac{ L }{ L^{\mathrm{coh}}_{kj} }
\right)^2
-
2 \pi^2 \xi^2
\left(
\frac{ \sigma_x }{ L^{\mathrm{osc}}_{kj} }
\right)^2
\right]
\,,
\label{017}
\end{equation}
with
the oscillation and coherence lengths
\begin{eqnarray}
&&
L^{\mathrm{osc}}_{kj}
=
\frac{ 4 \pi E }{ \Delta{m}^2_{kj} }
\,,
\label{0181}
\\
&&
L^{\mathrm{coh}}_{kj}
=
\frac{ 4 \sqrt{2} E^2 }{ |\Delta{m}^2_{kj}| }
\sigma_x
\,,
\label{0182}
\end{eqnarray}
and
\begin{eqnarray}
&&
\sigma_x^2
=
{\sigma_x^{\mathrm{P}}}^2
+
{\sigma_x^{\mathrm{D}}}^2
\,,
\label{0191}
\\
&&
\xi^2 \sigma_x^2
=
\xi_{\mathrm{P}}^2 {\sigma_x^{\mathrm{P}}}^2
+
\xi_{\mathrm{D}}^2 {\sigma_x^{\mathrm{D}}}^2
\,.
\label{0192}
\end{eqnarray}
This result was already obtained\footnote{
The only improvement in the expressions above with respect
to the analogous ones in Ref.~\cite{Giunti-Kim-Coherence-98}
is the introduction of $\xi_{\mathrm{D}}$,
that takes into account the properties of the detection process.
}
in a quantum mechanical framework
in Ref.~\cite{Giunti-Kim-Coherence-98},
where the wave packet treatment with pure states presented
several years before in Ref.~\cite{Giunti-Kim-Lee-Whendo-91}
was extended in order
to take into account the coherence properties of the
detection process,
whose importance was first recognized in Ref.~\cite{Kiers:1996zj}.
Expressions for the
$\nu_\alpha\to\nu_\beta$
transition probability similar
to Eq.~(\ref{017})
have been also obtained with wave packet treatments
of neutrino oscillations
in a quantum field theoretical framework
in Refs.~\cite{Giunti-Kim-Lee-Lee-93,%
Giunti-Kim-Lee-Whendo-98,%
Cardall-Coherence-99,%
Beuthe:2002ej,%
hep-ph/0205014}.

It is important to remark that
the wave packet treatment of neutrino oscillations
confirms the standard value in Eq.~(\ref{0181})
for the oscillation length
$L^{\mathrm{osc}}_{kj}$.
The coherence length
$L^{\mathrm{coh}}_{kj}$
in Eq.~(\ref{0182})
is the distance beyond which
the interference of the massive neutrinos
$\nu_k$ and $\nu_j$
is suppressed because
the separation of their wave packets when they arrive at the detector
is so large that they cannot be absorbed coherently.
As shown by Eqs.~(\ref{0182}) and (\ref{0191}) the
coherence lengths
$L^{\mathrm{coh}}_{kj}$
are proportional to the total coherence size,
that is dominated by the largest between the coherence sizes
$\sigma_x^{\mathrm{P}}$
and
$\sigma_x^{\mathrm{D}}$
of the production and detection process.
The last term in the exponential of Eq.~(\ref{017})
implies that the interference of the massive neutrinos
$\nu_k$ and $\nu_j$
is observable only if the
localization of the production and detection processes
is smaller than the oscillation length.

The localization term is important
for the distinction of
neutrino oscillation experiments
from experiments on the measurement of neutrino masses.
As first shown by Kayser in Ref.~\cite{Kayser:1981ye},
neutrino oscillations are suppressed
in experiments able to measure the value of a neutrino mass,
because the measurement of a neutrino mass
implies that only the corresponding massive neutrino is
produced or detected.

Kayser's \cite{Kayser:1981ye}
argument goes as follows.
Since a neutrino mass
is measured from energy-momentum conservation
in a process in which a neutrino is produced or detected,
from the energy-momentum dispersion relation (\ref{005}),
the uncertainty of the mass determination is
\begin{equation}
\delta{m_k}^2
=
\sqrt{ \left( 2 E_k \delta{E_k} \right)^2 + \left( 2 p_k \delta{p_k} \right)^2 }
\simeq
2 \sqrt{2} E \sigma_p
\,,
\label{022}
\end{equation}
where the approximation holds for realistic extremely relativistic neutrinos
and
$\sigma_p$
is the momentum uncertainty.
If the mass of $\nu_k$ is measured
with an accuracy better than
$\Delta{m}^2_{kj}$,
\textit{i.e.}
\begin{equation}
\delta{m_k}^2
<
|\Delta{m}^2_{kj}|
\quad
\Longleftrightarrow
\quad
\frac{ |\Delta{m}^2_{kj}| }{ 2 \sqrt{2} E \sigma_p }
>
1
\,,
\label{023}
\end{equation}
the neutrino $\nu_j$ is not produced or detected\footnote{
The smallness of $\sigma_p$
required to satisfy the condition (\ref{023})
implies that
$\nu_k$ and $\nu_j$
cannot be produced or detected in the same process,
because
their momentum difference is larger than the momentum uncertainty,
or
their energy difference is larger than the energy uncertainty
(for extremely relativistic neutrinos the energy uncertainty
is practically equal to the momentum uncertainty
$\sigma_p$).
In this case only one of the two massive neutrinos
is produced or detected.
}
and the interference of
$\nu_k$ and $\nu_j$
is not observed.

The localization term
in the oscillation probability
(\ref{017})
automatically implements Kayser's mechanism.
Indeed,
it can be written as
\begin{equation}
2 \pi^2 \xi^2
\left(
\frac{ \sigma_x }{ L^{\mathrm{osc}}_{kj} }
\right)^2
=
\xi^2
\left(
\frac{ \Delta{m}^2_{kj} }{ 4 \sqrt{2} E \sigma_p }
\right)^2
\,,
\label{024}
\end{equation}
with the momentum uncertainty
\begin{equation}
\frac{1}{\sigma_p^2}
=
4 \sigma_x^2
=
\frac{1}{ {\sigma_p^{\mathrm{P}}}^2 }
+
\frac{1}{ {\sigma_p^{\mathrm{D}}}^2 }
\,.
\label{025}
\end{equation}
If the condition (\ref{023}) for neutrino mass measurement
is satisfied,
the localization term (\ref{024})
suppresses\footnote{
This argument assumes that $\xi$ is not too small.
As discussed in Ref.~\cite{hep-ph/0205014},
in realistic cases $\xi$ is a number of order one,
but,
at least in principle,
there could be some cases in which
$\xi$ is very small, or even zero.
In these cases Kayser's mechanism cannot be implemented in the simplified
quantum mechanical framework presented here.
However,
it has been shown in Ref.~\cite{Beuthe:2002ej}
that a wave packet quantum field theoretical derivation of the neutrino oscillation probability
produces additional terms, independent from $\xi$ or similar quantities,
which suppress oscillations when
the condition (\ref{023}) is satisfied.
I would like to thank M. Beuthe for a illuminating discussion
about this problem.
}
the interference of
$\nu_k$ and $\nu_j$.

In the above derivation of neutrino oscillations
the wave-packet description of massive neutrinos is crucial in order to
allow the integration over time of
the time-dependent density matrix operator $\hat\rho_\alpha(x,t)$
in Eq.~(\ref{012}),
which leads to the time-independent density matrix operator $\hat\rho_\alpha(x)$
in Eq.~(\ref{013}).
Of course,
as emphasized in Ref.~\cite{Stodolsky-unnecessary-98},
\begin{quote}
A single, given density matrix
can arise in different ways,
especially when incoherence is involved.
\end{quote}
As already noticed in Ref.~\cite{Kiers:1996zj}
the density matrix operator $\hat\rho_\alpha(x)$
can also be generated through an appropriate
incoherent average over the energy spectrum.
The same is obviously true for the
oscillation probability (\ref{017}).

\begin{figure}[p]
\begin{center}
\includegraphics*[bb=81 427 424 749, height=7cm]{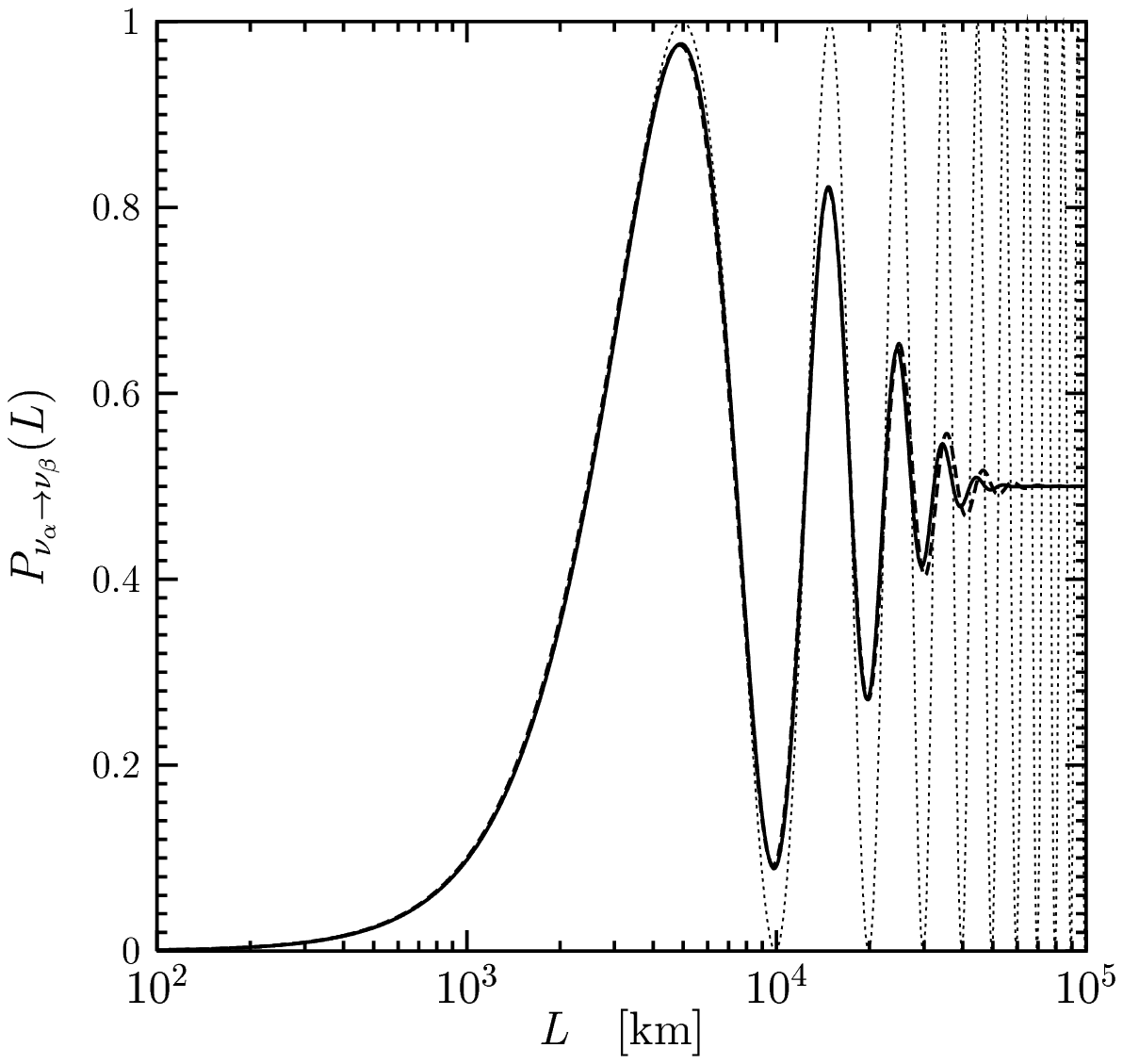}
\end{center}
\caption{ \label{prob}
Oscillation probability
as a function of distance $L$
for
$ \Delta{m}^2 = \unit{2.5 \times 10^{-3}}{\squaren{\electronvolt}} $,
$ \sin^2 2\vartheta = 1 $,
and
$ E = \unit{10}{\giga\electronvolt} $.
Dotted line:
Unsuppressed and unaveraged oscillation probability.
Dashed line:
Oscillation probability (\ref{035})
averaged over a gaussian
energy spectrum with width
$ \sigma_E = \unit{1}{\giga\electronvolt} $.
Solid line:
Oscillation probability (\ref{036})
suppressed by decoherence
due to a momentum uncertainty
$ \sigma_p = \unit{1}{\giga\electronvolt} $.
}
\end{figure}

In order to illustrate this point, let us consider
the simplest case of two-neutrino mixing.
The wave packet oscillation probability (\ref{017}) becomes
\begin{equation}
P_{\nu_\alpha\to\nu_\beta}(L)
=
\frac{1}{2}
\sin^2 2\vartheta
\left\{
1
-
\cos \left(
\frac{ \Delta{m}^2 L }{ 2 E }
\right)
\exp\left[
-
\left(
\frac{ L }{ L^{\mathrm{coh}} }
\right)^2
-
2 \pi^2 \xi^2
\left(
\frac{ \sigma_x }{ L^{\mathrm{osc}} }
\right)^2
\right]
\right\}
\,,
\label{036}
\end{equation}
with
the oscillation and coherence lengths
$L^{\mathrm{osc}}$ and $L^{\mathrm{coh}}$
given, respectively, by Eqs.~(\ref{0181}) and (\ref{0182})
with
$ \Delta{m}^2_{kj} = \Delta{m}^2 $
($\Delta{m}^2$ is the squared-mass difference
and
$\vartheta$ is the mixing angle;
see the reviews in Refs.~\cite{Bilenkii:2001yh,%
Gonzalez-Garcia:2002dz,%
Kayser:2002qs,%
Bilenky:2002aw}).
An incoherent average
of the probability in the plane-wave approximation
over a gaussian energy spectrum with width
$\sigma_E$
reads
\begin{equation}
P_{\nu_\alpha\to\nu_\beta}^{\mathrm{incoh}}(L)
=
\frac{1}{2}
\sin^2 2\vartheta
\left\{
1
-
\int \frac{ \mathrm{d}E' }{ \sqrt{ 2 \pi } \sigma_E }
\cos \left(
\frac{ \Delta{m}^2 L }{ 2 E' }
\right)
\exp\left[
-
\left(
\frac{ \left( E - E' \right)^2 }{ 2 \sigma_E^2 }
\right)^2
\right]
\right\}
\,.
\label{035}
\end{equation}
Figure~\ref{prob}
shows the values of the probabilities
(\ref{035}) and (\ref{036})
as functions of the source-detector distance $L$
for
$ \Delta{m}^2 = \unit{2.5 \times 10^{-3}}{\squaren{\electronvolt}} $,
$ \sin^2 2\vartheta = 1 $,
$ E = \unit{10}{\giga\electronvolt} $,
and
$ \sigma_E = \sigma_p = \unit{1}{\giga\electronvolt} $.
One can see that it is hard to distinguish the
dashed line obtained with Eq.~(\ref{035})
from the
solid line obtained with Eq.~(\ref{036}).

This result does not mean that wave packet effects are irrelevant\footnote{
Let us clarify here that with
``wave packet effects''
we mean effects due to the momentum and energy uncertainty
of a single process associated with
the interruption of the emitted wave train
caused by decay (natural linewidth)
or collisions (collision broadening)
discussed in Section~\ref{No source of waves vibrates indefinitely}.
These effects are usually not taken into account
in the calculation of neutrino energy spectra
(see, for example, Ref.~\cite{Bahcall:1994cf}).
}.
It means that in practice one can calculate with reasonable approximation the
decoherence of oscillations due to wave packet effects
either taking into account the momentum spread in the calculation of the amplitude,
as done in the derivation of Eq.~(\ref{036}),
or averaging the probability
over the same energy spread,
as done in the derivation of Eq.~(\ref{035})\footnote{
One must only be careful to notice that
the energy distribution in the incoherent average
of the probability must be normalized to one,
whereas the squared modulus of the momentum distribution of the wave packets
must be normalized to one.
}.
As emphasized in Ref.~\cite{Kiers:1996zj},
it also means that
if one does not have control on both the production and detection processes,
one cannot know if
oscillations are suppressed because of incoherent averaging over energy
of different microscopic processes
or because of decoherence due to the separation of wave packets.
However,
having good control of both the production and detection processes,
it may be possible in the far future to
reduce the causes of incoherent broadening of the energy spectrum
and
prove experimentally that
oscillations can be suppressed because of wave packet effects
(in practice using an approach similar to the
one adopted for the measurement of
the natural linewidth and collision broadening of atomic lines
discussed in Section~\ref{No source of waves vibrates indefinitely}).

Let us finally mention that
the fact that
coherent and incoherent
stationary beams
are indistinguishable
without a theoretical analysis
is well known in optics
(see Section 7.5.8 of Ref.~\cite{Born-Wolf-PrinciplesOfOptics-1959}),
in neutron interferometry
\cite{Comsa-PRL51-1105-1983,Kaiser-Werner-George-PRL51-1106-1983}
and
in general stationary particle beams
\cite{Bernstein-Low-PRL59-951-1987}.

\section{Energy of massive neutrinos}
\label{Energy of massive neutrinos}

In the recent Ref.~\cite{Lipkin:2002sq}
the claim
that different massive neutrinos
must have the same energy
\cite{Lipkin-nonexperiments-95,Grossman-Lipkin-spatial-97},
which was refuted in
Refs.~\cite{Giunti:2000kw,Giunti:2001kj},
has been renewed.
If such claim were correct,
it would mean that, at least in a quantum mechanical framework,
a wave packet description of neutrinos is not necessary.
Indeed,
in this case one can construct a time-independent
density matrix operator from the plane wave state
\begin{equation}
|\tilde\nu_\alpha(x,t)\rangle
=
\sum_k U_{\alpha k}^* e^{i p_k x - i E t} |\nu_k\rangle
\,,
\label{201}
\end{equation}
with
$p_k = \sqrt{ E^2 - m_k^2 }$:
\begin{equation}
\hat{\tilde\rho}_\alpha(x)
=
|\tilde\nu_\alpha(x,t)\rangle
\langle\tilde\nu_\alpha(x,t)|
=
\sum_{k,j}
U_{\alpha k}^* U_{\alpha j}
\exp\left[
i \left( p_k - p_j \right) x
\right]
\,.
\label{211}
\end{equation}
Using this density matrix operator
one can easily derive the standard oscillation probability
without need of neutrino wave packets.
Therefore,
in this section we consider recent claims
in favor of an equal energy of different massive neutrinos
and show that they are faulty.

The author of Ref.~\cite{Lipkin:2002sq} wrote:
\begin{quote}
\ldots
states with different ENERGIES ARE NEVER COHERENT
in any realistic experiment.
States of the same energy and different momenta
can be coherent,
but may not be.
\end{quote}
The reason offered by the author of Ref.~\cite{Lipkin:2002sq} is
\begin{quote}
The usual detector is a nucleon,
which changes its state after absorbing a neutrino and emitting
a charged lepton,
and is initially either in an energy eigenstate
or in a statistical mixture
in thermal equilibrium with its surroundings.
No neutrino detector has ever been
prepared in a coherent mixture
of energy eigenstates and no such
detector has been proposed
for future experiments.
\end{quote}
First,
one must note that in terminology of Ref.~\cite{Lipkin:2002sq}
``detector''
is not a macroscopic device,
but a nucleon hit by a neutrino.
Then,
the claim in Ref.~\cite{Lipkin:2002sq} is easily proved to be wrong
using the arguments presented
in Section~\ref{No source of waves vibrates indefinitely},
where we have shown that all kinds of waves,
classical or quantistic,
can be emitted or absorbed
only as wave packets,
\textit{i.e.}
as coherent superpositions
of plane waves with different frequency (energy).
If the claim in Ref.~\cite{Lipkin:2002sq} were correct,
photons, as well as other particles,
could never be emitted or absorbed.
The claim that a nucleon should be initially in an energy eigenstate
has been shown to be incorrect in Section~\ref{Bound states}.
Furthermore,
it is a mystery why all the claims on energy in Ref.~\cite{Lipkin:2002sq},
if they were true,
do not apply also to momentum.

Also the author of
Ref.~\cite{Stodolsky-unnecessary-98}
claimed that
different massive neutrinos have the same energy:
\begin{quote}
\emph{Energy or momentum?}
In mixing problems,
where we have to deal with linear combinations of particles
of different mass,
the question comes up as to whether one should
deal with states of the same energy or the same momentum.
Since as stated above,
for stationary conditions
we are to perform the calculation as an incoherent sum over energies,
we have given the answer ``energy''.
Evidently,
for stationary problems
it is most natural to use stationary wave functions
$\sim e^{-iEt}$.
\end{quote}
The meaning of this paragraph
seems to be that
different massive neutrinos are supposed to manage to have equal
energy in order to satisfy the
calculational needs of some theoretical physicists.
Therefore,
we cannot consider it a proof of anything.

A simple way to show the absurdity of the claims
that massive neutrinos must have equal energies
is to consider the transformation properties
of energy and momentum from one inertial frame to another,
following the discussion presented in Ref.~\cite{Giunti:2001kj}.

If the arguments presented in
Ref.~\cite{Lipkin:2002sq}
were correct,
in order to be produced and detected coherently
massive neutrinos should necessarily have exactly equal energies
in the inertial reference frames of both the
production and detection processes
(which presumably coincide with the rest frames of the initial nucleons in the
production and detection processes,
in the approach of Ref.~\cite{Lipkin:2002sq};
let us skip the useless discussion of the determination of the relevant inertial reference frame
in neutrino-electron scattering).
Such a requirement would imply that the
production and detection processes
have to be at rest in exactly the same inertial system,
because,
as shown in Ref.~\cite{Giunti:2001kj}
the equal-energy requirement is not Lorentz invariant.

Therefore
if the arguments presented in
Ref.~\cite{Lipkin:2002sq}
were correct it would mean that in practice
neutrino oscillations are not observable,
since it is practically impossible to
suppress random thermal motion in the source and detector.
Certainly,
oscillations of solar and atmospheric neutrinos
would be impossible.
Let me remind that in this discussion
we are concerned with the oscillatory terms in the
flavor transition probability
(\ref{017}),
due to the interference of different massive neutrinos,
which require coherence.
If these terms are suppressed,
it is still possible to measure the distance-independent
and energy-independent flavor-changing
probability
\begin{equation}
P_{\nu_\alpha\to\nu_\beta}
=
\sum_{k}
|U_{\alpha k}|^2 |U_{\beta k}|^2
\,,
\label{031}
\end{equation}
or the distance-independent flavor-changing
probability calculated in Ref.~\cite{Parke:1986jy}
if matter effects are important
\cite{Wolfenstein:1978ue,Mikheev:1985gs,Mikheev:1986wj}.
According to the results of solar neutrino experiments
\cite{Cleveland:1998nv,%
Hampel:1998xg,%
Altmann:2000ft,%
astro-ph/0204245,%
Fukuda:2002pe,%
Ahmad:2002jz,%
Ahmad:2002ka},
the solar neutrino problem
is very likely due to
neutrino oscillations in matter,
in which the distance-independent flavor-changing
probability calculated in Ref.~\cite{Parke:1986jy}
is relevant.
However,
recently the KamLAND experiment
\cite{hep-ex/0212021}
observed
indications in favor of
distance and energy dependent oscillations in vacuum of reactor antineutrinos
due to the same mass-squared difference
responsible of solar neutrino oscillations.
Distance and energy dependent flavor-changing
transitions have been also observed
in the Super-Kamiokande
\cite{Fukuda:1998mi},
Soudan 2
\cite{Allison:1999ms}
and MACRO
\cite{Ambrosio:2000qy}
atmospheric neutrino experiments
and
in the long-baseline experiment K2K
\cite{Ahn:2002up}.
These experimental evidences
show without any doubt that the arguments presented in
Ref.~\cite{Lipkin:2002sq}
in favor of an equal energy for different massive neutrinos
are wrong.

Indeed, if we consider for example
atmospheric neutrinos
produced by decays in the atmosphere
of highly energetic pions and muons,
the rest frames of the source and detector are very different
and different massive neutrinos cannot have the same energy
in both frames.
If the arguments presented in
Ref.~\cite{Lipkin:2002sq}
were correct different massive neutrinos could not be
either produced or detected coherently,
in contradiction with the observed oscillations.
The same reasoning applies to reactor and accelerator neutrinos,
remembering that
according to the arguments presented in
Ref.~\cite{Lipkin:2002sq}
coherence is possible only if
the source and detector rest frames coincide \emph{exactly}
(which, by the way, is an absurd concept,
if velocity is a real-valued quantity).

The only rigorous and correct calculation which could
be erroneously interpreted\footnote{
Indeed,
I made such mistake in the first version of this paper appeared in the
\texttt{hep-ph} electronic archive.
I am deeply indebted to M. Beuthe for his comments that helped to correct this mistake
and my wrong criticisms of
Refs.~\cite{Beuthe:2002ej,Beuthe:2001rc}.
}
as a proof in favor of the equal energy assumption
has been presented in Ref.~\cite{Beuthe:2002ej}
(see also Section~2.5 of Ref.~\cite{Beuthe:2001rc}).
The argument can be illustrated with the
density matrix model of neutrino oscillations
presented in Section~\ref{Stationary beams}.
Instead of writing the density matrix operator (\ref{011})
in the form (\ref{012})
using the wave packets (\ref{009})
in which the integration over the momentum $p$
in Eq.~(\ref{004})
has been already performed,
one could write the density matrix operator (\ref{011})
using the wave functions (\ref{004}) as
\begin{equation}
\hat\rho_\alpha(x,t)
=
\sum_{k,j}
U_{\alpha k}^* U_{\alpha j}
\int \frac{ \mathrm{d}p \, \mathrm{d}p' }{ 2\pi }
\psi_k(p) \psi_j(p')
e^{i(p-p')x-i(E_k(p)-E_j(p'))t}
|\nu_k\rangle
\langle\nu_j|
\,.
\label{021}
\end{equation}
As in Section~\ref{Stationary beams},
the time-independent density matrix operator
that describes a stationary beam is given by the average
over time of $\hat\rho_\alpha(x,t)$.
The integral over $t$
of
$\exp[-i(E_k(p)-E_j(p'))t]$
yields a
$\delta(E_k(p)-E_j(p'))$,
which means that
``interference occurs only between wave packet components with the same energy''
\cite{Beuthe:2001rc}
(see Refs.\cite{Gabor-RMP28-260-1956,Sudarsky:1991gv}
for similar results in electron interferometry
and kaon oscillations).
This means that the time average of
$\hat\rho_\alpha(x,t)$
is equivalent to an appropriate incoherent average
over the contributions of the single-energy components
of the wave packets.
This is the reason why
coherent and incoherent
stationary beams cannot be distinguished
without a theoretical analysis,
as discussed in Section~\ref{Stationary beams}.

Let us emphasize, however,
that this argument does not mean that
different massive neutrinos have the same energy,
or that wave packets are unnecessary.
The wave packet nature of massive neutrinos
is necessary for the existence of components
with the same energy that produce the observable interference,
and
the average energies of the wave packets of
different massive neutrino
are in general different.

\section{Conclusions}
\label{Conclusions}

Starting from the well-known fact that
``no source of waves vibrates indefinitely'',
we have argued that
neutrinos, as all other particles, are naturally described by wave packets.
This is in agreement with the well known fact that
in quantum theory
localized particles are described by
wave packets.
Even when neutrinos are produced or detected
through interactions with particles in bound states,
they are described by wave packets,
because bound states in interaction with an external field
do not have a definite energy.

Since the production and detection processes
in neutrino oscillation experiments are localized in space-time,
a consistent description of neutrino oscillations
requires a wave packet treatment.
In particular,
the wave packet character of massive neutrinos
is crucial for
the derivation of neutrino oscillations in space,
because the group velocity establishes a connection between space and time
and allows the time average of the space and time dependent
probability (or density matrix).

Finally, we have shown that the claimed arguments in
favor of an equal energy of massive neutrinos
(which could allow a quantum mechanical derivation of neutrino oscillations
without wave packets)
are in contradiction with well known physical laws and phenomena.

\section*{Acknowledgements}
\label{Acknowledgements}

I would like to thank M. Beuthe and C. Cardall
for very interesting and useful comments
on the first version of this paper appeared in the
\texttt{hep-ph} electronic archive.

\end{document}